# Triple decomposition and sparse representation for noisy pressure-sensitive paint data

Koyo Kubota[1], Makoto Takagi[1], Tsubasa Ikami[2], Yasuhiro Egami[3], Hiroki Nagai[2], Takahiro Kashikawa[4], Koichi Kimura[4], Yu Matsuda[1, *]

1. Department of Modern Mechanical Engineering, Waseda University, 3-4-1 Ookubo, Shinjuku-ku, Tokyo, 169-8555, Japan

2. Institute of Fluid Science, Tohoku University, 2-1-1 Katahira, Aoba-ku, Sendai, Miyagi, 980-8577, Japan

3. Department of Mechanical Engineering, Aichi Institute of Technology, 1247 Yachigusa, Yakusa-Cho, Toyota, Aichi, 470-0392, Japan

4. Quantum Application Core Project, Quantum Laboratory, Fujitsu Research, Fujistu Ltd, Kawasaki, Kanagawa 211-8588, Japan

* corresponding author: Yu Matsuda

**Email:** y.matsuda@waseda.jp

**Abstract**
Triple decomposition is a useful analytical method for extracting the mean value, organized coherent motion, and stochastic part from a fluctuating quantity. Although the pressure-sensitive paint (PSP) method is widely used to measure the pressure distribution on a surface, the PSP data measuring near atmospheric pressure contain significant noise. Here, we perform triple decomposition of noisy PSP data. To construct phase-averaged data representing an organized coherent motion, we propose a relatively simple method based on a multi-dimensional scaling plot of the cosine similarity between each PSP datum. Then, the stochastic part is extracted by selecting phase-averaged data with an appropriate phase angle based on the similarity between the measurement and phase-averaged data. As a data-driven approach, we also reconstruct the pressure distribution based on the triple decomposition and the pressure data at sparse optimal sensor positions determined from the proper orthogonal decomposition modes of the stochastic part. The optimal sensor positions are determined as a combinatorial optimization problem and are estimated using Fujitsu computing as a service digital annealer. Based on the results obtained, the root mean square error between the pressure measured by a pressure transducer and the reconstructed pressure obtained by the proposed method is small, even when the number of modes and sensor points is small. The application of PSP measurement is expected to expand further, and the framework for calculating triple decomposition and sparse representation based on the decomposition will be useful for flow analysis.

**Introduction**
The Reynolds decomposition is known as a basic method for extracting the stochastic part of flow. In the Reynolds decomposition, a fluctuating quantity $f(\mathbf{x}, t)$ at position vector $\mathbf{x}$ and time $t$ is decomposed as
$$f(\mathbf{x}, t) = \bar{f}(\mathbf{x}) + f'(\mathbf{x}, t), \qquad (1)$$
where $\bar{f}(\mathbf{x})$ is the mean value for time and $f'(\mathbf{x}, t)$ is the deviation from the mean value (the stochastic part). The triple decomposition was proposed for the analysis of a flow field consisting of organized coherent motion. [1] In the triple decomposition, a fluctuating quantity $f(\mathbf{x}, t)$ at position vector $\mathbf{x}$ and time $t$ is decomposed as
$$f(\mathbf{x}, t) = \bar{f}(\mathbf{x}) + \tilde{f}(\mathbf{x}, \phi(t)) + f'(\mathbf{x}, t), \qquad (2)$$
where $\tilde{f}(\mathbf{x}, \phi(t))$ is the contribution of the organized coherent motion and $\phi(t)$ is the phase angle at time $t$. The decomposition is straightforwardly conducted when a reference signal is available. The phase angle is extracted from the reference signal, and $\tilde{f}(\mathbf{x}, \phi(t))$ is obtained. [1,2] For example, a raw signal is divided into multiple phase angle ranges based on a reference signal, and the signals in each phase angle range are averaged to construct a phase-averaged signal. [2] However, reference signals are generally not always available. As a method that does not require a reference signal, an optimal frequency-domain filter deduced from power-spectral estimates is used to extract $\tilde{f}(\mathbf{x}, \phi(t))$. [3] These methods are useful for data measured by a point measurement technique such as a hot-wire anemometer and laser doppler velocimeter (LDV). The proper orthogonal decomposition (POD) is used to extract a coherent structure for the analysis of spatial two-dimensional (2D) data such as particle image velocimetry (PIV). The extracted coherent structure reconstructed by truncated POD modes was similar to that obtained by the phase-averaged fields. [4] For example, the averaged pressure and acceleration structures are estimated from PIV data based on this method. [5] It has also been applied to the analysis of the coherent velocity fluctuation of swirling turbulent jets. [6] The triple decomposition based on the optimal mode decomposition (OMD), [7] which is a generalization of the dynamic mode decomposition (DMD), [8] is proposed to decompose a flow field consisting of large and small coherent structures. [9] This method is applied to the analysis of the effect of free-stream turbulence on the near-field growth of the near wake of a cylinder. [10]

In this study, we conduct the triple decomposition for noisy data as measured by the pressure-sensitive paint (PSP) method, [11-14] which is also known as a flow-field measurement method. The PSP method has been applied to various fields of aero dynamics such as aerospace engineering, [15-18] fluid machinery, [19-22] and automotive engineering. [23,24] The PSP method enables the measurement of the pressure on the surface to which the PSP coating is applied. The method utilizes oxygen quenching of photoluminescence, and the pressure is measured by the variation of the emission intensity of the PSP coating containing luminescent dye. Since the variation in the emission intensity is small for a small pressure variation, the detection of the small pressure variation is difficult due to the relatively large camera noise compared with the variation of the emission. Therefore, the PSP data contain significant noise. For PSP data, by explicitly describing the measurement noise $n(\mathbf{x}, t)$ at the position vector $\mathbf{x}$ and time $t$, we can decompose pressure $p(\mathbf{x}, t)$ as:

$$p(\mathbf{x}, t) = \bar{p}(\mathbf{x}) + \tilde{p}(\mathbf{x}, \phi(t)) + p'(\mathbf{x}, t) + n(\mathbf{x}, t), \tag{3}$$

where $\bar{p}(\mathbf{x})$ is the mean value for time, $\tilde{p}(\mathbf{x}, \phi(t))$ is the contribution of the organized coherent motion, $\phi(t)$ is the phase angle at time $t$, and $p'(\mathbf{x}, t)$ is the stochastic part. The difficulty with the decomposition for PSP data is to extract $\tilde{p}(\mathbf{x}, \phi(t))$ and $p'(\mathbf{x}, t)$ from the large measurement noise of $n(\mathbf{x}, t)$. The POD and DMD analysis can extract a large flow structure, and several methods using POD and DMD are proposed to reduce the noise from time-series PSP data. [25-30] These methods enable the selection of the modes to reconstruct the pressure distribution with reduced noise. For example, the empirical selection of POD modes with small noise [25] or mode selection by sparse modeling have been proposed. [28,30] Although these denoising methods effectively remove the noise, to the best of our knowledge, there are no reports on the application of POP and DMD to triple decomposition for PSP measurement data. This is because even the first few POD/DMD modes with large energy contributions contain the noise; [25,28,29] thus, it is difficult to separate a large-scale structure from the noise. Moreover, data representing a large flow structure reconstructed from these modes do not always coincide with phase-averaged data. As a phase-average method for the PSP method, the conditional image sampling (CIS) method is proposed; [31] the CIS method requires the acquisition of a reference signal. In our previous study, we proposed a phase-average method based on a time-series clustering method without using a reference signal. [32] Utilizing this phase-average method, we estimate $\tilde{p}(\mathbf{x}, \phi(t))$. Then, the stochastic part $p'(\mathbf{x}, t)$ is extracted from the noise $n(\mathbf{x}, t)$ based on the POD analysis based on sparse modeling. Based on the obtained stochastic part $p'(\mathbf{x}, t)$, we estimate optimal sensor positions to effectively represent the pressure distribution using Fujitsu computing as a service digital annealer. [33,34] Sparse sensing is realized at optimal sensor positions, which are determined to be the solution to the optimal sensor placement problem. [24,35-38] As a demonstration, we apply the method to the PSP data measuring the pressure distribution of the Kármán vortex in our previous studies. [39,40] The time-series pressure data is triple-decomposed and reconstructed as noise-suppressed data. The reconstructed pressure data are compared with those measured by a pressure transducer. This triple decomposition without using a reference signal is useful for analyzing flow field data with significant noise such as PSP data.

**Proposed Method for Triple Decomposition**

The pressure distribution $p(\mathbf{x}, t)$ measured by the PSP method can be decomposed into Eq. (3). First, the mean distribution for time is calculated. Second, the phase-averaged distribution is extracted as the coherent motion. Third, the stochastic part is obtained by calculating $p(\mathbf{x}, t) - \bar{p}(\mathbf{x}) - \tilde{p}(\mathbf{x}, \phi(t))$. Finally, the noise of the stochastic part data is reduced. The detail of the proposed method is provided in this section.

**Calculation of mean distribution for time**

Theoretically, the mean value of the pressure distribution $\bar{p}(\mathbf{x})$ is defined as

$$\bar{p}(\mathbf{x}) = \lim_{T \to \infty} \frac{1}{T} \int_0^T p(\mathbf{x}, t) \, \mathrm{d}t. \tag{4}$$

However, pressure $p(\mathbf{x}, t)$ is measured over a time interval of $\Delta t$ by a camera in the PSP method and the measurement time is always limited. The mean value of the measured pressure is calculated as follows:

$$\bar{p}(\mathbf{x}) = \frac{1}{N} \sum_{n=0}^{N} p(\mathbf{x}, t_n), \tag{5}$$

where $t_n$ is the measured time represented as $t_n = n\Delta t$ and $n = 0, 2, \cdots, N$. The total number of measurements (images for the PSP method) is represented by $N$.

**Estimation of coherent motion**

The pressure $\tilde{p}(\mathbf{x}, \phi(t))$ organized coherent motion is estimated by a phase-average method. In our previous study, we proposed a phase-average method that does not use a reference signal. [32] The brief explanation of the method is provided in this paper. The cosine similarities between each data points are calculated, and the cosine similarity $\cos \theta_{i,j}$ is defined as

$$\cos \theta_{i,j} = \frac{\langle p(\mathbf{x}, t_i), p(\mathbf{x}, t_j) \rangle}{\|p(\mathbf{x}, t_i)\|_2 \|p(\mathbf{x}, t_j)\|_2}, \tag{6}$$

where $\langle p(\mathbf{x}, t_i), p(\mathbf{x}, t_j) \rangle$ is the inner product of $p(\mathbf{x}, t_i)$ and $p(\mathbf{x}, t_j)$, and $\|p(\mathbf{x}, t_i)\|_2$ is the $\ell 2$ norm of $p(\mathbf{x}, t_i)$. The integers $i$ and $j$ are within the range of $0 \leq i \leq N$ and $0 \leq j \leq N$, respectively. Since the pressure $p(\mathbf{x}, t_i)$ are usually a 2D array, $p(\mathbf{x}, t_i)$ is reshaped into a column vector in this similarity calculation. Then, the measured pressure data are presented on a 2D scatter plot using multi-dimensional scaling (MDS) based on the distance metric $|\sin(\theta_{i,j}/2)|$ between $p(\mathbf{x}, t_i)$ and $p(\mathbf{x}, t_j)$, where $|\sin(\theta_{i,j}/2)|$ is the absolute value of $\sin(\theta_{i,j}/2)$. The data on the circle with a radius of $1/2$ in the MDS plot represent a periodic phenomenon. On the other hand, the data not on the circle are out of the periodic one. In our previous study, the data were classified as a combinatorial optimization problem using Fujitsu computing as a service digital annealer. As a simplified method, we manually classify the data for each phase angle in this study. Then, the phase-averaged data $\tilde{p}_M(\mathbf{x}, \phi_m)$ are obtained by averaging the data in each phase angle range. The subscript $M$ denotes the number of divisions per one period. When one period is divided into $M$ subperiods, the range of each phase angle is $2\pi/M$ and $m$-th phase angle $\phi_m$, which is defined as the center of each range, is written as

$$\phi_m = (2m - 1)\frac{\pi}{M}, \tag{7}$$

where $m$ is the integer within the range of $1 \leq m \leq M$.

**Estimation of stochastic part**

The sum of the stochastic part $p'(\mathbf{x}, t_n)$ and noise $n(\mathbf{x}, t)$ at time $t_n = n\Delta t$ is calculated as

$$p'(\mathbf{x}, t_n) + n(\mathbf{x}, t_n) = p(\mathbf{x}, t_n) - \bar{p}(\mathbf{x}) - \tilde{p}(\mathbf{x}, \phi(t_n)). \tag{8}$$

The problem is to determine the phase angle $\phi(t_n)$ because a reference signal is not used in this study. Moreover, the phase-averaged data $\tilde{p}_M(\mathbf{x}, \phi_m)$ are obtained at the discrete phase angles of $\phi_m$. Following the POD/OMD approach proposed by Baj et al., [9] which focus on flow velocity not pressure, $\tilde{p}(\mathbf{x}, \phi(t_n))$ is estimated by the minimization problem as follows:

$$\tilde{p}(\mathbf{x}, \phi(t_n)) = \underset{\tilde{p}(\mathbf{x}, \phi(t_n))}{\mathrm{argmin}} \|p(\mathbf{x}, t_n) - \bar{p}(\mathbf{x}) - \tilde{p}(\mathbf{x}, \phi(t_n))\|_2, \tag{9}$$

$$\tilde{p}(\mathbf{x}, \phi(t_n)) = \sum_k c_i(t_n) \psi_k(\mathbf{x}), \tag{10}$$

where $\psi_k(\mathbf{x})$ is the $k$-th POD/OMD mode and $c_i(t_n)$ is the coefficient at time $t_n$ corresponding to the mode $\psi_k(\mathbf{x})$. However, this least-squares method will result in overfitting for noisy data. That is, this method will estimate a part of $p'$ as a part of $\tilde{p}$.

In this study, we propose an alternative method. We extend the discrete phase-averaged data using the following method: First, $\tilde{p}_M(\mathbf{x}, \phi_m)$ is reshaped to a column vector, which is written as $\tilde{\mathbf{p}}_M(\phi_m)$. Second, the matrix $\widetilde{\mathbf{P}}_M$ with the column vectors arranged side-by-side is considered as follows:

$$\widetilde{\mathbf{P}}_M = [\,\tilde{\mathbf{p}}_M(\phi_1) \;\; \tilde{\mathbf{p}}_M(\phi_2) \;\; \cdots \;\; \tilde{\mathbf{p}}_M(\phi_M)\,]. \tag{11}$$

The singular value decomposition (SVD) of $\widetilde{\mathbf{P}}$ is written as

$$\widetilde{\mathbf{P}}_M = \mathbf{U}\mathbf{\Sigma}\mathbf{V}^\top, \tag{12}$$

where $\mathbf{U}$ and $\mathbf{V}$ are unitary matrices and the superscript T indicates the transpose. The matrix $\mathbf{\Sigma}$ is a diagonal matrix whose diagonal elements are singular values. The columns of $\mathbf{U}$ are left singular vectors that correspond to the POD modes. The columns of $\mathbf{V}$ are right singular vectors representing the phase history of each left singular vector. Third, fitting the amplitude variations $\mathbf{V}^\top$ with a curve, the estimated data with an increased number of divisions $L$ ($M < L$) per one period can be obtained. In other words, we can estimate super-resolution data in the phase direction of the phase-averaged data as

$$\widetilde{\mathbf{P}}_L = \mathbf{U}\mathbf{\Gamma}, \tag{13}$$

where $\mathbf{\Gamma}$ is the matrix representing the amplitudes determined by the singular value $\mathbf{\Sigma}$ and the interpolated values by curve fitting. The number of the divisions $L$ is determined by calculating the average of the residuals between $\widetilde{\mathbf{p}}_L(\phi_l)$ and the next phase range $\widetilde{\mathbf{p}}_L(\phi_{l+1})$, where $l$ is the integer within the range of $1 \leq l \leq L$. Then, the average of the residuals $\mathcal{R}$ is calculated as

$$\mathcal{R} = \frac{1}{qL}\left\|\widetilde{\mathbf{P}}_{L+1} - \widetilde{\mathbf{P}}_L\right\|_F^2, \tag{14}$$

$$\widetilde{\mathbf{P}}_L = [\ \widetilde{\mathbf{p}}_L(\phi_1)\ \ \widetilde{\mathbf{p}}_L(\phi_2)\ \ \cdots\ \ \widetilde{\mathbf{p}}_L(\phi_L)\ ], \tag{15}$$

$$\widetilde{\mathbf{P}}_{L+1} = [\ \widetilde{\mathbf{p}}_L(\phi_L)\ \ \widetilde{\mathbf{p}}_L(\phi_1)\ \ \cdots\ \ \widetilde{\mathbf{p}}_L(\phi_{L-1})\ ], \tag{16}$$

where $q$ is the number of pixels of the image and the subscript $F$ indicates the Frobenius norm. It is expected that $\mathcal{R}$ will converge for a sufficiently large $L$. Finally, $\widetilde{p}(\mathbf{x}, \phi(t_n))$ is selected as $\widetilde{p}_L(\mathbf{x}, \phi_l)$, which is equivalent to $\widetilde{\mathbf{p}}_L(\phi_l)$ in the reshaped vector form, with the highest cosine similarity to $p(\mathbf{x}, t_n) - \bar{p}(\mathbf{x})$.

The stochastic part $p'(\mathbf{x}, t_n)$ is separated from the noise $n(\mathbf{x}, t_n)$. The sum of the stochastic part $p'(\mathbf{x}, t_n)$ and noise $n(\mathbf{x}, t)$ is denoted as

$$\hat{p}(\mathbf{x}, t_n) = p'(\mathbf{x}, t_n) + n(\mathbf{x}, t_n). \tag{17}$$

Here, we again use SVD. The sum $\hat{p}(\mathbf{x}, t_n)$ is reshaped to a column vector $\hat{\mathbf{p}}(t_n)$, and the matrix $\widehat{\mathbf{P}} = [\hat{\mathbf{p}}(0)\ \hat{\mathbf{p}}(\Delta t)\ \hat{\mathbf{p}}(2\Delta t)\ \cdots\ \hat{\mathbf{p}}(n\Delta t)\ \cdots\ \hat{\mathbf{p}}(N\Delta t)]$ is considered. The SVD of $\widehat{\mathbf{P}}$ is written as

$$\widehat{\mathbf{P}} = \widehat{\mathbf{U}}\widehat{\mathbf{\Sigma}}\widehat{\mathbf{V}}^\top, \tag{18}$$

where $\widehat{\mathbf{U}}$ and $\widehat{\mathbf{V}}$ are unitary matrices and $\mathbf{\Sigma}$ represents singular values. The stochastic part $\mathbf{p}'(t_n)$, which is the reshaped form of $p'(\mathbf{x}, t_n)$, is extracted by truncating the SVD; thus, the stochastic part $\mathbf{p}'(t_n)$ is written as

$$\mathbf{p}'(t_n) = \widehat{\mathbf{U}}\boldsymbol{\alpha}(t_n), \tag{19}$$

where $\boldsymbol{\alpha}(t_n)$ is the amplitude vector. The amplitude $\boldsymbol{\alpha}(t_n)$ is estimated using sparse modeling based on the least absolute shrinkage and selection operator (LASSO). [41-43] However, the estimation of $\boldsymbol{\alpha}(t_n)$ for all pixels of all measured images results in high computational cost. In our previous studies, [30,44] we proposed an estimation method for the pressure distribution $p(\mathbf{x}, t_n)$ based on the selected optimal sensor points determined as the optimal sensor placement problem. Here, we propose the optimal points for detecting the stochastic part $\mathbf{p}'(t_n)$. The same algorithm using Fujitsu computing as a service digital annealer [44] can be applied, except that the pressure distribution is replaced by the stochastic part $\mathbf{p}'(t_n)$. We briefly introduce the algorithm for determining optimal points. More details are provided in our previous study. [44] The idea of the method is that the large variation of the POD mode at an optimal point well represents the characteristic of the distribution and points with similar variation of the mode do not need to be selected. We consider the row vector $\hat{\mathbf{u}}_i$ of $\widehat{\mathbf{U}}\widehat{\mathbf{\Sigma}}$, and $\hat{\mathbf{u}}_i$ represents the variation of the mode at point $i$. The weight $w(\hat{\mathbf{u}}_i, \hat{\mathbf{u}}_j)$ between points $i$ and $j$ is defined as

$$w(\hat{\mathbf{u}}_i, \hat{\mathbf{u}}_j) = |\hat{\mathbf{u}}_i||\hat{\mathbf{u}}_j||\sin\theta|, \tag{20}$$

where $\theta$ is the angle between $\hat{\mathbf{u}}_i$ and $\hat{\mathbf{u}}_j$. The optimal sensor points are selected by solving the maximum clique problem from the undirected graph $\mathcal{G}_c$, whose vertices have the weights larger than $c$. We solve the independent set problem of the complement graph $\bar{\mathcal{G}}_c$ of $\mathcal{G}_c$, because a maximum clique problem is equivalent to an independent set problem of the complement graph in general. This problem is known as an NP-hard problem and is solved by Fujitsu computing as a service digital annealer. The independent set is obtained by solving the following minimization problem:

$$\min - \sum_{\zeta \in \bar{\mathcal{G}}_c} x_\zeta + \Lambda \sum_{(\zeta,\eta) \in \Omega} x_\zeta x_\eta \,, \qquad (21)$$

where $x_\zeta = 0$ or $1$ for the $\zeta$-th vertex belonging to or not to $\bar{\mathcal{G}}_c$, respectively. The set $\Omega$ represents the edge set of $\bar{\mathcal{G}}_c$. Then, the amplitude $\boldsymbol{\alpha}(t_n)$ is determined by the following minimization problem of LASSO [30]

$$\boldsymbol{\alpha}(t_n) = \underset{\boldsymbol{\alpha}(t_n)}{\mathrm{argmin}} \frac{1}{2q} \|\mathbf{S}\hat{\mathbf{p}}(t_n) - \mathbf{S}\hat{\mathbf{U}}\boldsymbol{\alpha}(t_n)\|_2^2 + \lambda \|\boldsymbol{\alpha}(t_n)\|_1 \,, \qquad (22)$$

where $\|\boldsymbol{\alpha}(t_n)\|_1$ is the $\ell 1$ norm of $\boldsymbol{\alpha}(t_n)$ and $\mathbf{S}$ is the matrix indicating optimal sensor positions. The element of $\mathbf{S}$ is 1 for the sensor position and 0 for other positions. Then, the stochastic part $\mathbf{p}'(t_n)$ is estimated by Eq. (19) using the calculated $\boldsymbol{\alpha}(t_n)$.

## Results and Discussion
### Triple mode decomposition for PSP data

We applied the proposed method to the PSP data obtained in our previous study. [39,40] The data was the pressure distribution induced by the Kármán vortex behind a square cylinder. The Reynolds number was $1.1 \times 10^5$. The time interval of $\Delta t$ was $1.0$ ms. Before applying the proposed method, the PSP distributions were processed by a $5 \times 5$ smoothing spatial filter. A typical example of the processed data is shown in Fig. 1. As shown in the figure, the noise was still large for further analysis.

Next, we calculated the cosine similarity for the 21800 data points by Eq. (6). Since the data contained large amounts of noise, a truncated-SVD of rank five was used to reduce the noise. It is noted that these truncated-SVD data were only used for calculation of the cosine similarity not for the following phase-averaging calculation. The MDS plot is shown in Fig. 2. The data on the circle with a radius of $1/2$ shown in the red dashed line represents a periodic phenomenon; thus, we obtained the phase-averaged data by dividing the data on the circle in 12 subsets ($M = 12$) and averaging each subset. Here, the number of the divisions per period $M = 12$ was determined because the amount of data was not enough to reduce the noise by averaging for some phase range for larger $M$. The phase-averaged data are shown in Fig. 3, where the phase angle $\phi_m$ ($m = 1, 2, \cdots, 12$) is defined in Eq. (7). In this figure, the phase-averaged distributions of $\bar{p}(\mathbf{x}) + \tilde{p}_{M=12}(\mathbf{x}, \phi_m)$ are presented for ease of visibility. The noise was significantly reduced by averaging.

Based on these discreate phase averaged data, the super-resolution data in the phase direction of the phase-averaged data were estimated. The SVD of the phase-averaged data was calculated as Eq. (12). As an example, the first six POD modes (column vectors of $\mathbf{U}$) are shown in Fig. 4. We employed the first four modes for estimating the super-resolution data based on the threshold determination method proposed by Donoho and Gavish. [45] The amplitudes for the 1st and 3rd POD modes are shown in Fig. 5. The discrete amplitudes for phase angle were interpolated by following function:

$$v_h(\phi) = \sum_{k=1}^{2} a_k \sin(b_k \phi + c_k) \,, \qquad (23)$$

where $v_h(\phi)$ is the amplitude for $h$-th mode at phase $\phi$. For the first mode, $a_1 = 0.41$, $b_1 = 1.00$, $c_1 = 0.61$, $a_2 = 0.01$, $b_2 = 3.00$, and $c_2 = -1.00$. The first mode contains the fundamental and third harmonics. However, the effect of the third harmonic is small. For the 3rd mode, $a_1 = -0.01$, $b_1 = 0.97$, $c_1 = -1.11$, $a_2 = -0.41$, $b_2 = 2.00$, and $c_2 = -2.10$. The third mode contains the fundamental and second harmonics. Then, the number of divisions per period was determined to be $L = 96$ based on Eq. (14), where $\mathcal{R}$ converged almost to $2.0 \times 10^{-8}$. The, the super-resolution data were estimated by Eq. (13). Then, $\tilde{p}(\mathbf{x}, \phi(t_n))$ was selected as $\tilde{p}_L(\mathbf{x}, \phi_l)$ with the maximum cosine similarity to $p(\mathbf{x}, t_n) - \bar{p}(\mathbf{x})$. Here, the amplitude of $\tilde{p}_L(\mathbf{x}, \phi_l)$ was also determined by maximizing the similarity because the amplitude of the vortex varied. Figure 6 shows the selected phase for $p(\mathbf{x}, t_n) - \bar{p}(\mathbf{x})$ over time up to 128 ms. As shown in the figure, the selected phase-averaged image changed periodically over time. Figures 7 and 8 show typical images of $\hat{p}(\mathbf{x}, t_n)$ shown in Eq. (17) and the first six POD modes of $\hat{p}(\mathbf{x}, t_n)$, respectively. While the significant noise was observed on $\hat{p}(\mathbf{x}, t_n)$, the noise on the POD modes is small due to the SVD calculation based on a large number of images (21800 images). Based on the POD modes, we calculated optimal sensor positions by Eq. (21) and show an example of sensor positions in Fig. 9. The optimal sensors were placed where the vortices passed. This trend was similar to the results calculated based on the POD mode of the Kármán vortex in previous study. [44] Then, the stochastic part $p'(\mathbf{x}, t_n)$ was extracted from $\hat{p}(\mathbf{x}, t_n)$ by Eq. (22) as shown in Fig. 10. Finally, the noise reduced pressure distribution was obtained by calculating the sum $\bar{p}(\mathbf{x}) + \tilde{p}(\mathbf{x}, \phi(t_n)) + p'(\mathbf{x}, t_n)$ as shown in Fig. 11, and the pressure distributions were successfully reconstructed by the method. As shown at $t_6$ in Fig. 11, twin vortices were sometimes observed. In such case, the similarity was small for any $\tilde{p}_L(\mathbf{x}, \phi_l)$ and the amplitude of $\tilde{p}_L(\mathbf{x}, \phi_l)$ was also small. This correlation between the similarity and amplitude was clearly observed as shown in Fig. 12. This indicates that the contribution of $\tilde{p}_L(\mathbf{x}, \phi_l)$ is small for the phenomena that deviate from the periodic coherent motion.

**Sparse sensing at optimal sensor points**
As a data-driven approach, the pressure distribution can be reconstructed from the pressure data at the optimal sensor points estimated by the POD modes of $\hat{p}(\mathbf{x}, t_n)$ based on triple decomposition. The mean value $\bar{p}(\mathbf{x})$, organized coherent motion $\tilde{p}_L(\mathbf{x}, \phi_l)$, and optimal sensor positions $\mathbf{S}$ were given/pre-calculated as a prior knowledge in a data-driven approach. The reconstruction of the pressure distribution was conducted by the following method. First, we calculated the cosine similarities between the pressure data at optimal sensor positions and $\tilde{p}_L(\mathbf{S}\mathbf{x}, \phi_l)$ ($l = 1, 2, \cdots, L$) and selected $\tilde{p}_L(\mathbf{x}, \phi_{l_{\max}})$ at the phase $\phi_{l_{\max}}$ that gives the maximum similarity. Second, $\hat{p}(\mathbf{x}, t_n)$ was calculated using $\bar{p}(\mathbf{x})$ and $\tilde{p}_L(\mathbf{x}, \phi_{l_{\max}})$. Third, the SVD of $\hat{p}(\mathbf{x}, t_n)$ was calculated and the amplitudes of the POD modes were determined following Eq. (22) to extract $p'(\mathbf{x}, t_n)$ using Eq. (19) from the noisy data of $\hat{p}(\mathbf{x}, t_n)$. Finally, the pressure distributions were obtained by $\bar{p}(\mathbf{x}) + \tilde{p}_L(\mathbf{x}, \phi_{l_{\max}}) + p'(\mathbf{x}, t_n)$.

The root mean square error (RMSE) between the pressure data measured by a pressure transducer through a pressure tap and the reconstructed data are shown in Fig. 13. The position of the pressure tap is shown in Fig. 13(a). For comparison, the RMSE results obtained by the method based on the previous study [30,44] were also shown. The pressure distributions were reconstructed from the POD modes calculated from the observed data $p(\mathbf{x}, t)$, and the amplitudes of these POD modes were determined by LASSO at the optimal sensor positions that were determined by these POD modes in the previous method. That is, the difference between these methods is whether the sensor position is determined based on the POD modes of $\hat{p}(\mathbf{x}, t_n)$. The horizontal axis in Fig. 13(b) represents the number of POD modes used to reconstruct the pressure distribution, and this was varied by varying $\lambda$ in LASSO. The RMSE was calculated for 128 time steps and the standard deviation of the number of modes used for each time step is shown in the figure. The results for the number of optimal sensor points of 25 and 60 are shown. The RMSE values were small for a small number of optimal sensor points and POD modes in the proposed method. This result indicates that triple decomposition enables a sparser representation.

**Conclusions**
We conducted the triple decomposition of noisy PSP data. We propose a relatively simple method based on the MDS plot of the cosine similarity to construct phase-averaged data representing an organized coherent motion for periodic flows. Since the constructed phase-averaged data are typically discrete with respect to the phase angle direction, we produced the data with a sufficient phase angle step size by interpolating the data between the phase angles. The stochastic part can be extracted by selecting phase-averaged data with an appropriate phase angle based on the similarity between the measurement data and phase-averaged data. The noise of the stochastic part was suppressed by superposing the POD modes whose amplitudes were determined by LASSO at optimal sensor positions. We used Fujitsu computing as a service digital annealer for the estimation of optimal sensor positions based on the POD modes of the stochastic part. In this manner, the triple decomposition of the noisy PSP data was performed. The application of PSP measurements is expected to expand further, and triple decomposition will be very useful.

We also reconstructed the pressure distribution from the pressure data at optimal sensor positions as a data-driven approach. A sparser representation was achieved using phase-averaged data, the POD modes of the stochastic part, and the optimal sensor positions determined from these POD modes. Moreover, the number of sensors required for reconstruction was reduced compared with that in the previous study.

**Data availability**
The part of flow measurement dataset is available in zenodo with the following identifier:
https://doi.org/10.5281/zenodo.10215642.

**Acknowledgments**

A part of this work was supported by the Collaborative Research Project J23I041, J24I064 and the Low Turbulence Wind Tunnel Facility of the Advance Flow Experimental Research Center at the Institute of Fluid Science, Tohoku University. The authors would like to express their gratitude to Dr. Yasuhumi Konishi, Mr. Hiroyuki Okuizumi and Mr. Yuya Yamazaki for their assistance during the wind tunnel testing. We also gratefully appreciate Tayca Corporation for providing the titanium dioxide.

**Author Contributions**

Koyo Kubota: Conceptualization, Data curation, Formal analysis, Investigation, Methodology, Software, Validation, Visualization.

Makoto Takagi: Investigation, Methodology, Software

Tsubasa Ikami: Data curation, Investigation, Resources, Visualization, Writing -review & editing

Yasuhiro Egami: Data curation, Formal analysis, Investigation, Resources, Writing -review & editing

Hiroki Nagai: Investigation, Resources, Writing -review & editing

Takahiro Kashikawa: Methodology, Software, Validation

Koichi Kimura: Methodology, Software, Validation

Yu Matsuda: Conceptualization, Data curation, Formal analysis, Funding acquisition, Investigation, Methodology, Project administration, Resources, Software, Supervision, Validation, Visualization, Writing – original draft, Writing – review & editing

**Competing Interest Statement**

The authors declare the following competing interests: Takahiro Kashikawa and Koichi Kimura are employees of Fujitsu Ltd. All other authors declare no competing interests.

**Corresponding Author**

Yu Matsuda; y.matsuda@waseda.jp

**Figures and Tables**

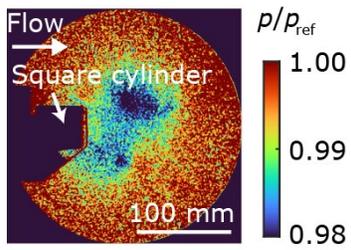

**Figure 1.** Typical pressure distribution data measured by PSP method, where pressure $p$ is normalized by an atmospheric pressure $p_{\text{ref}}$. A $5 \times 5$ smoothing spatial filter is applied to original data. Reproduced from [30].

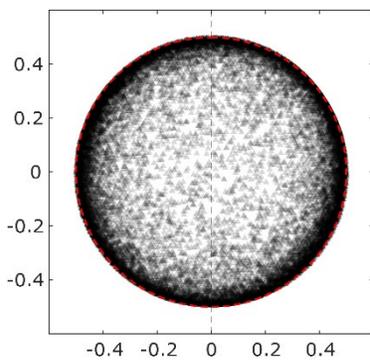

**Figure 2.** MSD plot based on cosine similarity between measurement data. The data on the circle with a radius of 1/2 shown in red dashed line represent a periodic phenomenon.

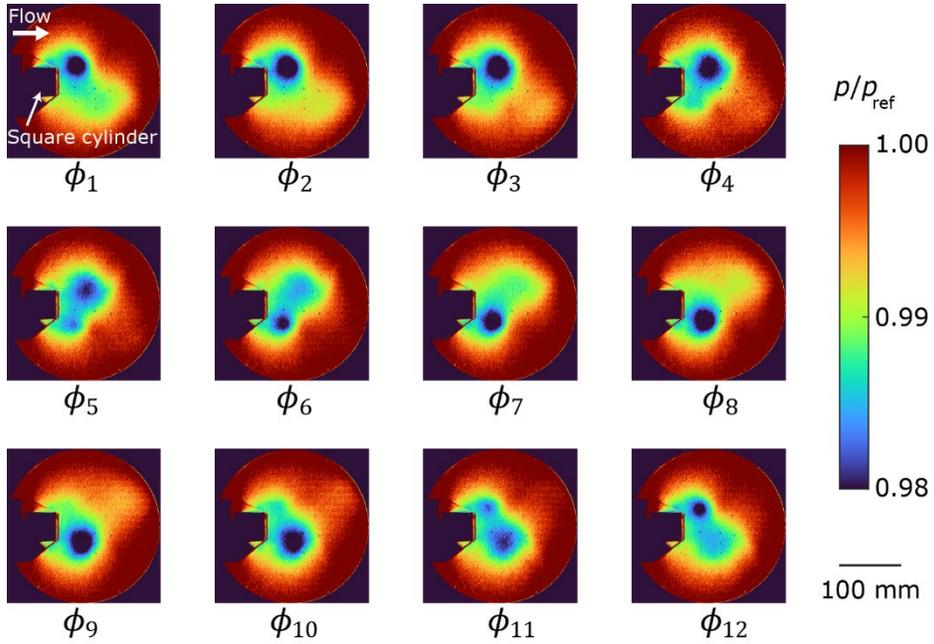

**Figure 3.** Phase averaged pressure data $\bar{p}(\mathbf{x}) + \tilde{p}_{M=12}(\mathbf{x}, \phi_m)$, where the pressure is normalized by the atmospheric pressure $p_{\text{ref}}$ and $\phi_n$ indicates the $n$-th phase.

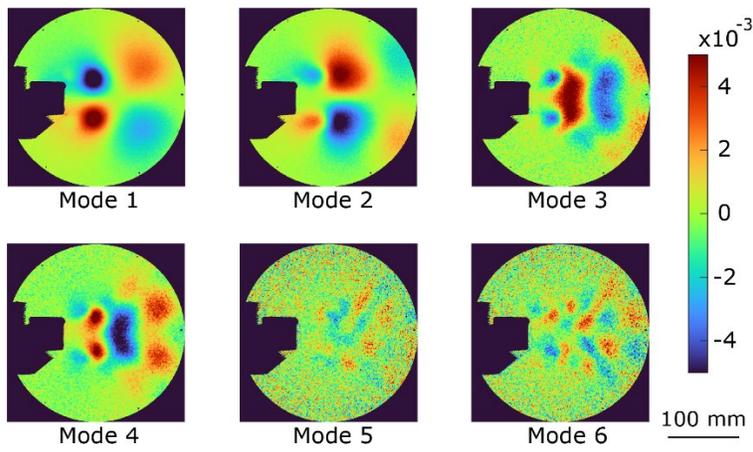

**Figure 4.** First six POD modes of phase averaged pressure data.

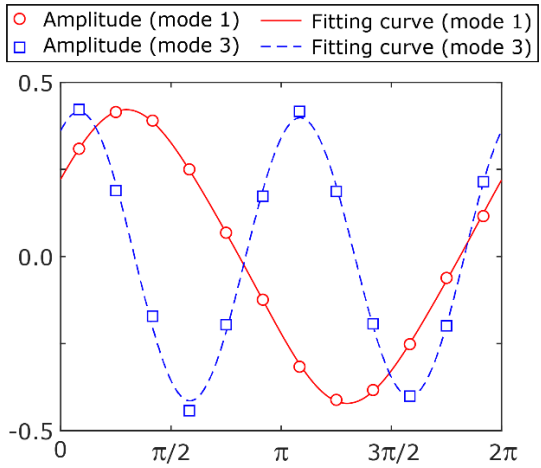

**Figure 5.** Typical example of amplitude of first and third POD modes. The horizontal and vertical axes are phase and amplitude, respectively. The fitting curves are used for interpolating the data between discrete phases.

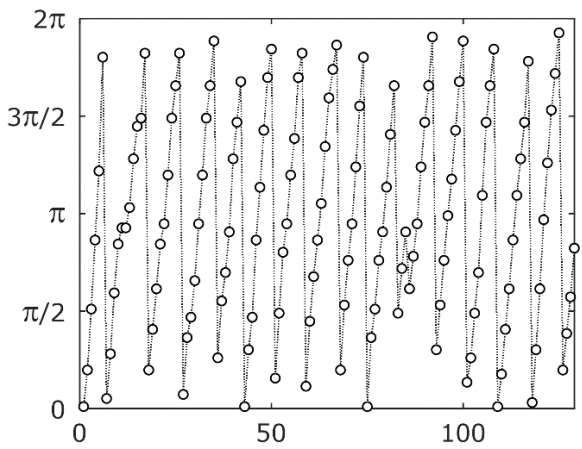

**Figure 6.** Typical example of selected phase for $p(\mathbf{x}, t_n) - \bar{p}(\mathbf{x})$ over time up to 128 ms. The horizontal and vertical axes are time and phase, respectively.

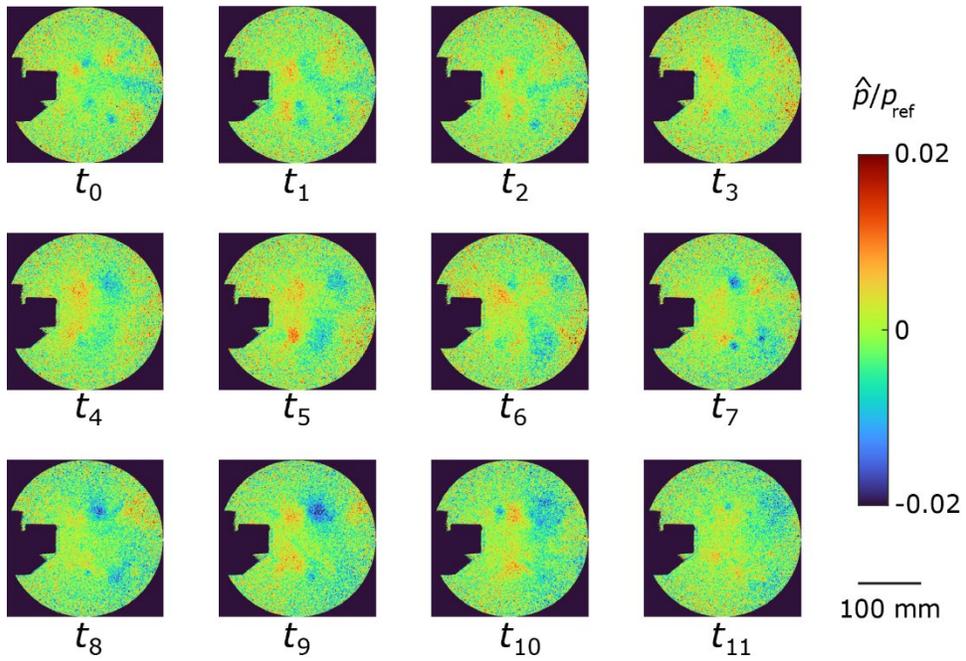

**Figure 7.** Typical example of extracted stochastic part containing noise $\hat{p}(\mathbf{x}, t_n)$, where $p_{\text{ref}}$ is the atmospheric pressure.

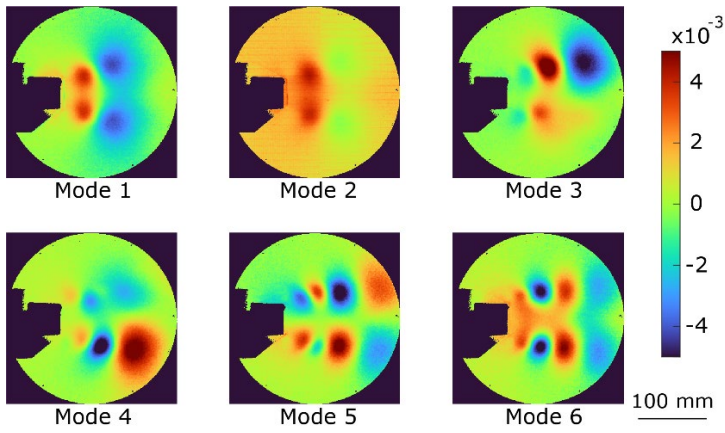

**Figure 8.** First six POD modes of stochastic part.

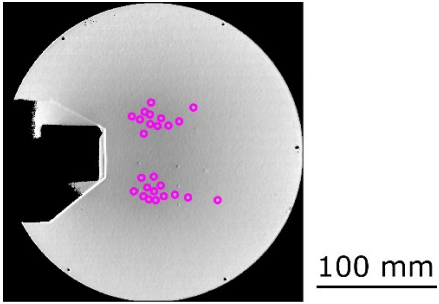

**Figure 9.** Example of optimal sensor positions calculated based on POD mode of stochastic part. The number of sensors is 25.

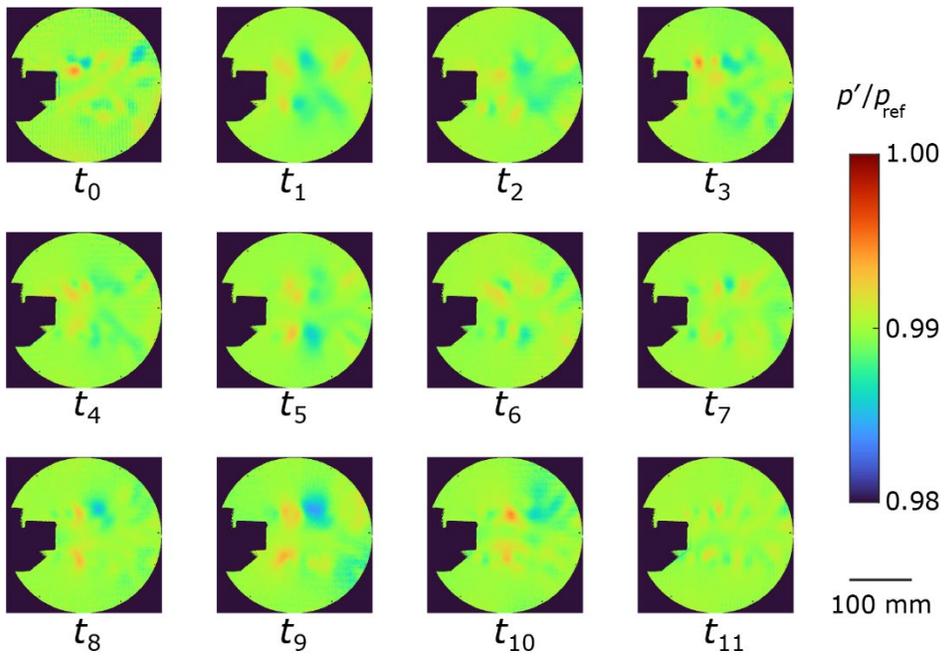

**Figure 10.** Typical example of denoised stochastic part $p'(\mathbf{x}, t_n)$, where $p_{\text{ref}}$ is the atmospheric pressure.

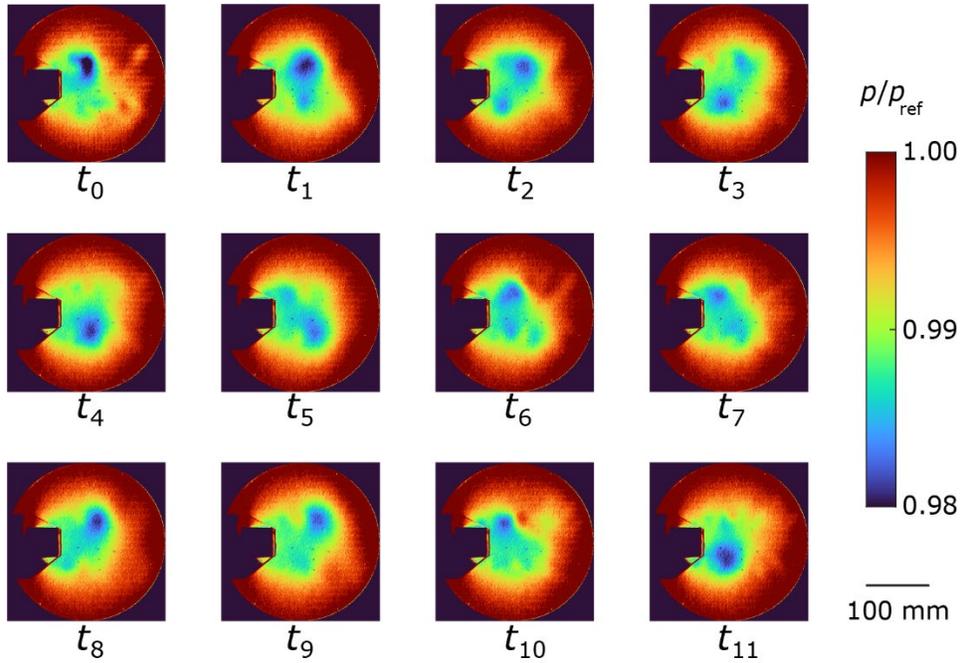

**Figure 11.** Typical example of denoised pressure distribution $p(\mathbf{x}, t_n)$, where $p_{\text{ref}}$ is the atmospheric pressure. The pressure $p(\mathbf{x}, t_n)$ is obtained by $\bar{p}(\mathbf{x}) + \tilde{p}_L(\mathbf{x}, \phi_l) + p'(\mathbf{x}, t_n)$.

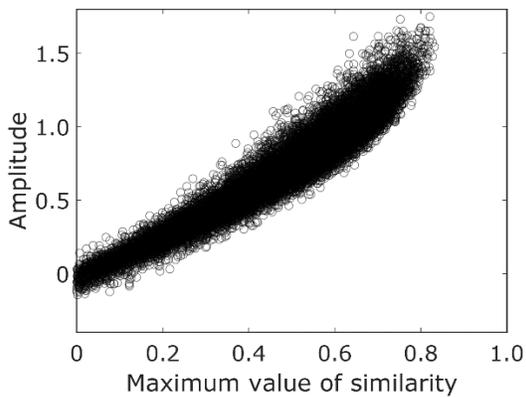

**Figure 12.** Relation between maximum value of similarity and amplitude. The amplitude is small for small similarity resulting in a small contribution of coherent motion $\tilde{p}_L(\mathbf{x}, \phi_l)$ to pressure distribution $p(\mathbf{x}, t_n)$.

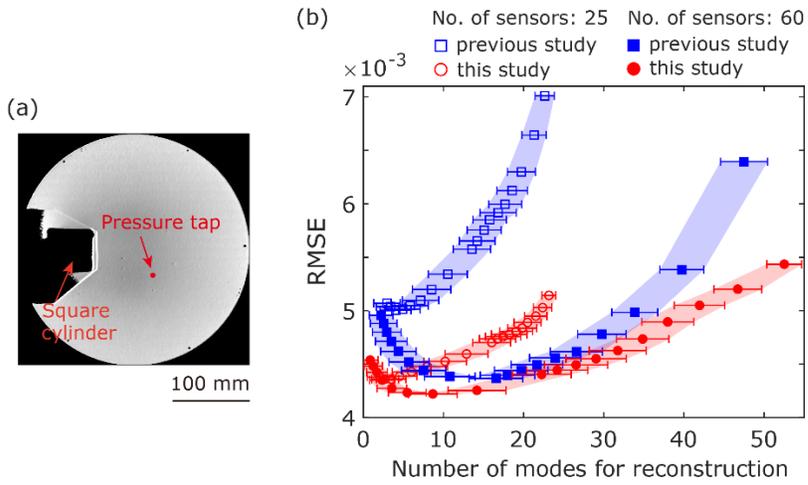

**Figure 13.** Comparison between proposed and previous methods for pressure reconstruction performance. (a) The position of the pressure tap. (b) The RMSE between reconstructed pressure and pressure measured at pressure tap. The horizontal axis is the number of POD modes used to reconstruct the pressure distribution.